\documentclass[draftcls, onecolumn]{IEEEtran}
\usepackage{cite}

\usepackage{latexsym,color,amsmath,amsthm,amssymb,amscd,amsfonts}
\usepackage{amsmath,amsfonts,amssymb}
\usepackage{fancyhdr,graphicx,amsmath, amscd}
\usepackage{color}

\def\be{\begin{equation}}
\def\ee{\end{equation}}
\def\bea{\begin{eqnarray}}
\def\eea{\end{eqnarray}}
\def\bt{\begin{theorem}}
\def\et{\end{theorem}}
\def\bl{\begin{lemma}}
\def\el{\end{lemma}}
\def\br{\begin{remark}}
\def\er{\end{remark}}
\def\bc{\begin{corollary}}
\def\ec{\end{corollary}}
\def\bd{\begin{definition}}
\def\ed{\end{definition}}

\def\b{\beta}

\def\l{\lambda}

\def\vP{\vec{P}}
\def\vQ{\vec{Q}}
\def\vf{\vec{f}}

\def\cK{\mathcal{K}}

\def\bP{\mathbf{P}}
\def\bQ{\mathbf{Q}}

\def\bbR{\mathbb{R}}

\def\b1{B_{1}^}

\def\dfpq{D_{\vec{\mathbf{f}}} (\vec{\bP}, \vec{\bQ})}
\def\ba{\begin{array}}
\def\ea{\end{array}}
\def\ben{\begin{enumerate}}
\def\een{\end{enumerate}}
\newtheorem{theorem}{Theorem}[section]
\newtheorem{lemma}{Lemma}[section]
\newtheorem{remark}{Remark}[section]
\newtheorem{proposition}{Proposition}[section]
\newtheorem{corollary}{Corollary}[section]

\newtheorem{definition}{Definition}[section]
\def\dfpqi{D_{\vf}(\vP, \vQ; i)}
\def\bff{\mathbf{f}}

\begin{document}
%
\title{On the mixed $f$-divergence for multiple pairs of measures}  
%
%
%

\author{Elisabeth~ M.~ Werner,~
                and~Deping~Ye
\thanks{E.M. Werner  is with the Department of Mathematics, Case Western Reserve University, Cleveland, OH, USA, 44106, and Universit\'{e} de Lille 1, UFR de Math\'{e}matique, 59655 Villeneuve d'Ascq, France. email: elisabeth.werner@case.edu} 
\thanks{D. Ye is with the Department of Mathematics and Statistics, Memorial University of Newfoundland, St. John's, Newfoundland, Canada A1C 5S7. email: deping.ye@mun.ca}
}

\maketitle

\begin{abstract}
In this paper,  the concept of the classical $f$-divergence (for a pair of measures) is extended to the mixed $f$-divergence (for multiple pairs of measures). The mixed $f$-divergence provides a way to measure the difference between multiple pairs of (probability) measures. Properties for the mixed $f$-divergence are established, such as permutation invariance and symmetry in distributions. An
Alexandrov-Fenchel type inequality and an isoperimetric type inequality for the 
mixed $f$-divergence will be proved and applications in the theory of convex bodies are given.

\end{abstract}


%
\IEEEpeerreviewmaketitle

\section{Introduction} 

 \IEEEPARstart{I}{n  applications} such as pattern matching, image analysis, statistical learning, and information theory, one often needs to compare two (probability) measures and to know whether they are similar to each other. Hence, finding the ``right" quantity to measure the difference between two (probability) measures $P$ and $Q$ is central. Traditionally, people use  classical $L_p$ distances between $P$ and $Q$, such as the variational distance and/or the $L_2$ distance. However,  the family of $f$-divergences is often more suitable to fulfil the goal than the classical $L_p$ distance of measures. 
\par
The $f$-divergence $D_f(P, Q)$ of two probability measures $P$ and $Q$ was first introduced in \cite{Csiszar}, and independently in \cite{AliSilvery1966, Morimoto1963} as
 \begin{equation}
 D_f(P, Q)=\int_Xf\left(\frac{p}{q}\right) q\,d\mu. \label{f:divergence}
\end{equation}
Here,  $p$ and $q$ are  density functions of  $P$ and $Q$ with respect to  a measure $\mu$ on $X$.
The idea behind the $f$-divergence is to replace, for instance, the function $f(t)=|t-1|$ in the variational distance by a general convex function $f$. Hence the $f$-divergence includes various widely used divergences as special cases, such as, the variational distance, the  Kullback-Leibler  divergence \cite{KullbackLeibler1951}, the Bhattcharyya distance \cite{Bhattacharyya1946} and many more. Consequently,  the $f$-divergence receives considerable attention  
(e.g.,  \cite{BarronGyorfiMeulen, CoverThomas, HarremoesTopsoe, LieseVajda2006, OsterrVajda}). We also refer to,  for instance \cite{Basseville2010},  for more references related to the $f$-divergence. \par
Extension of the $f$-divergence from two (probability) measures to multiple (probability) measures is fundamental in many applications,  such as statistical hypothesis test and classification, and much  research has been devoted to that, for instance in   \cite{  Menendez, MoralesPardo1998, Zografos1998}. Such extensions include, e.g., the Matusita's affinity \cite{Matusita1967, Matusita1971}, the Toussaint's affinity \cite{Toussaint1974},  the information radius \cite{Sibson1969} and the average divergence \cite{Sgarro1981}. 
\par
 The $\bf f$-dissimilarity $D_{\bff}(P_1, \cdots, P_l)$ for (probability) measures $P_1, \cdots, P_l$,  introduced in \cite{GyorfiNemetz1975, GyorfiNemetz1978} for a convex function  $\bff : \bbR^l\rightarrow \bbR$, is   
 a natural generalization of the $f$-divergence. 
 It is defined as
 \begin{equation*} 
D_\bff(P_1, \cdots, P_l)=\int_X \bff(p_1, \cdots, p_l)\,d\mu, \end{equation*}  where the $p_i$'s are density functions of the $P_i$'s that are absolutely continuous with respect to $\mu$. 
For a convex function $f$, the function $\bff(x,y)=y f(\frac{x}{y})$ is also convex on $x, y>0$, and $D_{\bff}(P, Q)$ is equal to the classical $f$-divergence defined in formula (\ref{f:divergence}). Note that the Matuista's affinity is related to  $$\bff(x_1,\cdots, x_l)=-\prod_{i=1}^lx_i^{1/l},$$ and the Toussaint's affinity is related to  $$\bff(x_1,\cdots, x_l)=-\prod_{i=1}^lx_i^{a_i}, \ \ a_i\geq 0\ \ with\ \  \sum_{i=1}^la_i=1.$$ 

Inspired by the growing fascinating  connections between convex geometry and information theory
(e.g., \cite {JenkinsonWerner, LutwakYangZhang2002/1, 
LutwakYangZhang2004/1, LutwakYangZhang2005, PaourisWerner2011, Werner2012/1, Werner2012b}), we introduce  special   $\bf f$-dissimilarities, namely the mixed $f$-divergence and the $i$-th mixed $f$-divergence. These will be done in Section 2 and Section 5 of this paper. 
Also in Section 2, we establish some basic properties of the mixed $f$-divergence.
In Section 3, we focus on the $\bff$-dissimilarity and the mixed $f$-divergence for multiple convex bodies. In particular, we show that the general mixed affine surface area -- a fundamental concept in convex geometry -- is a special case of the mixed $f$-divergence. An Alexandrov-Fenchel type inequality and an isoperimetric type inequality for the mixed $f$-divergence are obtained in Section 4. Section 5 is dedicated to the $i$-mixed $f$-divergence and its related isoperimetric  type inequalities.

\section{The Mixed $f$-Divergence.}

Throughout this paper, let $(X, \mu)$ be a finite measure space. For $1 \leq i
\leq n$, let  $ P_i=p_i  \mu$ and  $ Q_i=q_i \mu$ be probability
measures on $X$ that are absolutely continuous with respect to
the measure $\mu$. {Moreover, we assume that for all $i=1,  \cdots, n$,  $p_i$ and $q_i$ are nonzero  almost everywhere w.r.t. the measure $\mu$.} We use $\vec{\bP}$ and  $\vec{\bQ}$ to denote the vectors of probability measures, or, in short,   probability vectors,  $$\vec{\bP}=(P_1, P_2, \cdots, P_n), \ \ \ \vec{\bQ}=(Q_1, Q_2, \cdots, Q_n).$$ We  use $\vec{p}$ and $\vec{q}$ to denote the vectors of density functions, or density vectors,  for $\vec{\bP}$ and $\vec{\bQ}$ respectively, $$\frac{\,d\vec{\bP}}{\,d\mu}= \vec{p}=(p_1, p_2, \cdots, p_n),\ \ \ \ \frac{\,d\vec{\bQ}}{\,d\mu} =\vec{q}=(q_1, q_2, \cdots, q_n).$$ 
We make the convention that $0 \cdot \infty =0$. 
 
Denote by $\mathbb{R}^+=\{x\in \mathbb{R}: x\geq 0\}$.  Let $f: (0, \infty) \rightarrow  \mathbb{R}^+$ be a non-negative convex or concave function. 
The $*$-adjoint function $f^*:(0, \infty) \rightarrow  \mathbb{R}^+$
of $f$  is defined by
\begin{equation*}
f^*(t) = t f (1/t).
\end{equation*}
It is obvious that $(f^*)^*=f$ and that $f^*$ is again convex,  respectively concave, if
$f$ is convex,  respectively concave.
 
\par
 Let $f_i: (0, \infty) \rightarrow
\mathbb{R}^+$, $ 1 \leq i \leq n$, be either convex or concave functions. Denote by $\vec{\mathbf{f}} =(f_1, f_2, \cdots, f_n)$ the vector of functions. We write $$\vec{\mathbf{f}}^*=(f_1^*, f_2^*, \cdots, f_n^*)$$ to be the $*$-adjoint vector for $\vec{\mathbf{f}}$. 
\vskip 2mm 
Now  we introduce {\em the mixed $f$-divergence} for $(\vec{\mathbf{f}}, \vec{\bP}, \vec{\bQ})$  as follows. 
\vskip 2mm 
\begin{definition}\label{mixedf} Let  $(X, \mu)$ be a finite measure space. Let $\vec{\bP}$ and $\vec{\bQ}$ be two probability  vectors on $X$ with density vectors $\vec{p}$ and $\vec{q}$ respectively.  
The mixed $f$-divergence $\dfpq$ for $(\vec{\mathbf{f}}, \vec{\bP}, \vec{\bQ})$ 
is defined by
\begin{equation}\label{mixed1} 
\dfpq= \int_{X} \prod_{i=1}^n
\left[f_i\left(\frac{p_{i}}{q_{i}}\right) q_{i}\right]^\frac{1}{n}
d \mu.
\end{equation}
\end{definition}
\par
\noindent 
Similarly, we  define the mixed $f$-divergence for $(\vec{\mathbf{f}}, \vec{\bQ}, \vec{\bP})$ by
\begin{equation}\label{mixed2}
D_{\vec{\mathbf{f}}}(\vec{\bQ}, \vec{\bP})=  \int_{X} \prod_{i=1}^n
\left[f_i\left(\frac{q_{i}}{p_{i}}\right) p_{i}
\right]^\frac{1}{n} d\mu.
\end{equation}
A  special case is when all distributions $P_i$ and $Q_i$ are identical and equal to a probability distribution $P$. 
In this case, \begin{eqnarray*} \dfpq\!&=&\!\!\!D_{(f_1, f_2, \cdots, f_n)}\big((P, P, \cdots, P),(P, P, \cdots, P)\big)\\ \!&=&\!\!\! \prod_{i=1}^n\left[f_i(1)\right]^{\frac{1}{n}}.\end{eqnarray*} 

\vskip 2mm \noindent {\bf Remark.} The mixed $f$-divergence as defined in Definition \ref{mixedf}  is closely related to  the $\bff$-dissimilarity. In fact, taking $$\bff(x_1, y_1; \cdots; x_n, y_n)=-\prod_{i=1}^n\left\{y_if\left(\frac{x_i}{y_i}\right)\right\}^{\frac{1}{n}},$$ then  the $\bff$-dissimilarity is equal to the negative of the mixed $f$-divergence, namely, $$D_{\bff}(P_1, Q_1; \cdots; P_n, Q_n)=-\dfpq,$$ if $\bff$ is convex. In general, the function $\bff$ could be neither convex nor concave. However, if, for instance,  all $f_i$ are  twice differentiable concave functions, then $\bff$ is a convex function. Indeed, let  $\lambda\in [0,1]$, $x_1, x_2, y_1,  y_2 \in \bbR^+$ such that  $y_1\neq 0$ and $ y_2\neq 0$. Put  $x_{\lambda}=\lambda x_1+(1-\lambda)x_2$ and $y_{\lambda}=\lambda y_1+(1-\lambda)y_2$.  Then we have   for all $1\leq k\leq   n$, 
 \begin{eqnarray*}
  \left\{\!y_{\lambda} f_i\!\left(\!\frac{x_{\lambda}}{y_{\lambda}}\!\right)\!\right\}^{\frac{k}{n}}  
 \!\!\!\!\!\!\!\!&=&\!\!\!\! \left\{\!y_{\lambda}f_i\left(\frac{ x_1 }{ y_1} \cdot \frac{  \lambda y_1}{y_{\lambda} }\!+\!\frac{x_2}{y_2} \cdot \frac{  (1-\lambda) y_2}{y_{\lambda} }\!\right)\!\!\right\}^{\frac{k}{n}}\\ \!\!\!&\geq&\!\!\!\!  \left\{\!{\lambda}y_1f_i\left(\frac{ x_1 }{ y_1}\right)\!\!  +\!(1-\lambda)y_2f_i\left(\frac{x_2}{y_2} \right)\!\right\}^{\frac{k}{n}}\\ \!\!\!&\geq & \!\!\!\!  {\lambda}\! \left\{\!y_1 f_i\left(\frac{ x_1 }{ y_1}\right)\! \right\}^{\frac{k}{n}}\!\!\!\!+\!(1-\lambda)\left\{y_2  f_i\left(\frac{x_2}{y_2} \right)\!\!\right\}^{\frac{k}{n}} \end{eqnarray*} where the first inequality is from the concavity of $f_i$ and the monotone increasing of $t^{k/n}$; while the second inequality is from the concavity of $t^{k/n}$. 
 That is, the functions $[yf_i(x/y)]^{k/n}$ defined on $x, y\in \bbR^+$ are concave for all $1\leq i, k\leq n$. Therefore, the Hessian  of $\bff$ can be written as a block matrix with all diagonal matrices being positive semi-definite and all  off-diagonal blocks  equal to $0$. Consequently, the Hessian  of $\bff$ is positive semi-definite and hence $\bff$ is  convex.

\vskip 2mm
Let $\pi\in S_n$ denote a permutation on $\{1, 2, \cdots, n\}$ and denote $$\pi(\vec{p})=(p_{\pi(1)}, p_{\pi(2)},\cdots, p_{\pi(n)}).$$
One immediate result from  Definition \ref{mixedf} is the following permutation invariance for $\dfpq$. 
\par
\begin{proposition}
[\bf Permutation invariance] Let the vectors $\vec{\mathbf{f}}, \vec{\bP}, \vec{\bQ}$ be as above, and let $\pi\in S(n)$ be a permutation on $\{1, 2, \cdots, n\}$. Then  $$\dfpq=D_{\pi(\vec{\mathbf{f}})}(\pi(\vec{\bP}),\pi(\vec{\bQ})).$$ 
\end{proposition}
\vskip 2mm  
When all $(f_i, P_i, Q_i)$ are equal to
$(f, P, Q)$, the mixed $f$-divergence is equal to the classical $f$-divergence, denoted by $D_f(P, Q)$, which
takes the  form
\begin{eqnarray*} D_f(P, Q)&=&D_{(f, f, \cdots, f)}\big((P, P, \cdots, P), (Q, Q, \cdots, Q)\big)\\ &=&\int_{X}
f\left(\frac{p}{q}\right) q d \mu.\end{eqnarray*}
\vskip 2mm
As $f^*(t)=tf(1/t)$, one  easily obtains a fundamental property for the classical $f$-divergence $D_f(P, Q)$, namely,   $$D_f(P, Q)=D_{f^*}(Q, P),$$  for all $(f, P, Q)$. Similar results hold true for  the mixed $f$-divergence. We show this now.
\par 
Let $0\leq k\leq n$. We write $D_{\vec{\mathbf{f}}, k}(\vec{\bP}, \vec{\bQ})$ for 
\begin{eqnarray*}
D_{\vec{\mathbf{f}}, k}(\vec{\bP}, \vec{\bQ}) \!=\!\! \int_{X} \prod_{i=1}^k
\left[f_i\!\left(\!\frac{p_{i}}{q_{i}}\!\right)\! q_{i}\right]^\frac{1}{n}\!\!\times\!\! \prod_{i=k+1}^n\!
\left[f_i^*\!\left(\!\frac{q_{i}}{p_{i}}\!\right)\! p_{i}\right]^\frac{1}{n}\!\!
d \mu.\end{eqnarray*}
 Clearly, $D_{\vec{\mathbf{f}}, n}(\vec{\bP}, \vec{\bQ})=\dfpq$ and $D_{\vec{\mathbf{f}}, 0}(\vec{\bP}, \vec{\bQ})=D_{\vec{\mathbf{f}}^*}(\vec{\bQ}, \vec{\bP})$, where $$\vec{\mathbf{f}}^*=(f_1^*, f_2^*, \cdots, f_n^*).$$
Then we have the following result for  changing order of distributions.  
\begin{proposition}[\bf Principle for changing order of distributions] \label{principle}  Let $\vec{\mathbf{f}}, \vec{\bP}, \vec{\bQ}$ be as above.  Then, for any $0\leq k\leq n$, one has $$\dfpq=D_{\vec{\mathbf{f}}, k}(\vec{\bP}, \vec{\bQ}).$$
In particular, $$\dfpq=D_{\vec{\mathbf{f}}^*}(\vec{\bQ}, \vec{\bP}).$$ 
\end{proposition} 
\vskip 2mm \noindent \begin{IEEEproof} Let $0\leq k\leq n$. Then, \begin{eqnarray*}
\dfpq\!\!\!\!&=&\!\!\!\!\!\!\! \int_{X} \prod_{i=1}^k
\left[f_i\left(\!\frac{p_{i}}{q_{i}}\!\right)\! q_{i}\right]^\frac{1}{n}\!\!\!\times\!\!\! \prod_{i=k+1}^n\!\!
\left[f_i\left(\!\frac{p_{i}}{q_{i}}\!\right)\! q_{i}\right]^\frac{1}{n}
\!\!d \mu\\&=&\!\!\!\!\!\!   \int_{X} \prod_{i=1}^k
\left[f_i\left(\!\frac{p_{i}}{q_{i}}\!\right)\! q_{i}\right]^\frac{1}{n}\!\!\!\times\!\!\! \prod_{i=k+1}^n\!\!
\left[f_i^*\left(\!\frac{q_{i}}{p_{i}}\!\right) p_{i}\right]^\frac{1}{n}\!\!\!\!
d \mu \\ &=&\!\!\! D_{\vec{\mathbf{f}}, k}(\vec{\bP}, \vec{\bQ}),
\end{eqnarray*} where the second equality follows from $f_i\left(\frac{p_{i}}{q_{i}}\right) q_{i}=f_i^*\left(\frac{q_{i}}{p_{i}}\right) p_{i}$.   \end{IEEEproof}
\vskip 2mm
A direct consequence of Proposition \ref{principle} is the following symmetry principle for the mixed $f$-divergence. 

\begin{proposition} [\bf Symmetry in distributions] Let $\vec{\mathbf{f}}, \vec{\bP}, \vec{\bQ}$ be as above. Then,  $\dfpq+D_{\vec{\mathbf{f}}^*}(\vec{\bP}, \vec{\bQ})$ is symmetric in $\vec{\bP}$ and $\vec{\bQ}$, namely, $$\dfpq+D_{\vec{\mathbf{f}}^*}(\vec{\bP}, \vec{\bQ})=  D_{\vec{\mathbf{f}} }(\vec{\bQ}, \vec{\bP})+D_{\vec{\mathbf{f}}^*}(\vec{\bQ}, \vec{\bP}).$$
\end{proposition} 
\par
\noindent {\bf Remark.} Proposition \ref{principle} says that $\dfpq$ remains the same if one replaces any triple $(f_i, P_i, Q_i)$ by $(f_i^*, Q_i, P_i)$. It is also easy to see that, for all $0\leq k, l\leq n$, one has  $$\dfpq= D_{\vec{\mathbf{f}}, k}(\vec{\bP}, \vec{\bQ})=D_{\vec{\mathbf{f}}^*,l}(\vec{\bQ}, \vec{\bP})=D_{\vec{\mathbf{f}}^*}(\vec{\bQ}, \vec{\bP}).$$ Hence, for all $0\leq k, l\leq n$,  $$D_{\vec{\mathbf{f}}, k}(\vec{\bP}, \vec{\bQ})+D_{\vec{\mathbf{f}}^*,l}(\vec{\bP}, \vec{\bQ})=\dfpq+D_{\vec{\mathbf{f}}^*}(\vec{\bP}, \vec{\bQ})$$ is symmetric in $\vec{\bP}$ and $\vec{\bQ}$. 

\vskip 2mm
 Hereafter, we only consider the  mixed $f$-divergence $\dfpq$ defined in  formula (\ref{mixed1}).  Properties for the mixed $f$-divergence $D_{\vec{\mathbf{f}}}(\vec{\bQ}, \vec{\bP})$  defined in (\ref{mixed2})  follow along the same lines. 
 \par
Now we list  some important mixed $f$-divergences.

\vskip 2mm
\noindent
{\bf Examples.}
\par
\noindent (i) The total variation is a widely used $f$-divergence to measure the difference between two probability measures $P$ and $Q$ on $(X,\mu)$. It is related to function $f(t)=|t-1|$. Similarly, we  define the {\em mixed total variation} by $$D_{\vec{\mathbf{f}}}(\vec{\bP}, \vec{\bQ})=\int_X \prod_{i=1}^n |p_i-	q_i|^{\frac{1}{n}}\,d\mu.$$ It  measures the difference between two probability vectors $\vec{\bP}$ and $\vec{\bQ}$. 
\par \noindent
(ii) For $a \in \mathbb{R}$, we denote by $a_+=\max\{a,0\}.$ We define the {\em mixed relative entropy} or {\em mixed Kullback Leibler divergence} of  $\vec{\bP}$ and $\vec{\bQ}$ by
\begin{eqnarray*}
D_{KL}\big(\vec{\bP}, \vec{\bQ})\! =\! D_{(f_+, \cdots, f_+)}\big(\vec{\bP}, \vec{\bQ})\! = \!\! \int_{X} \prod_{i=1}^n
\left[ p_i \ln\left(\!\frac{q_{i}}{p_{i}}\!\right) \!
\right]_+^\frac{1}{n}\!\! d\mu,
\end{eqnarray*} where $f(t) = t \ln t$. 
When $ P_i=P=p \mu$ and $ Q_i= Q =q\mu$ for all $i=1, 2, \cdots, n$, we get the following  (modified) {\em relative entropy} or {\em  Kullback Leibler divergence} $$ D_{KL}\big(P || Q\big)= \int_{X} p \left[\ln\left(\frac{q}{p}\right)\right]_+ 
d\mu. $$
\par
\noindent
(iii) For the (convex and/or 
concave) functions $f_{\alpha_i}(t) = t^{\alpha_i}$, $\alpha_i \in \mathbb{R}$ for $1 \leq i \leq n$, we define the {\em mixed Hellinger integrals}
\begin{eqnarray*}
D_{  (f_{\alpha_1}, f_{\alpha_2}, \cdots, f_{\alpha_n}) }\big(\vec{\bP}, \vec{\bQ})=\int_{X} \prod_{i=1}^n
 \left[ p_i  ^\frac{\alpha_i}{n} q_i ^\frac{1-\alpha_i}{n}\right] 
 d\mu.
\end{eqnarray*}
In particular,  $$D_{  (t^{\alpha }, t^{\alpha }, \cdots, t^{\alpha }) }\big(\vec{\bP}, \vec{\bQ}) = \int_{X} \prod_{i=1}^n
 p_i  ^\frac{\alpha}{n} q_i ^\frac{1-\alpha}{n} 
 d\mu.$$ Those integrals are related to the Toussaint's affinity (see Introduction), and can be used to define the {\em mixed $\alpha$-R\'enyi divergence}
 \begin{eqnarray*}
D_{\alpha}\big(\{P_i || Q_i\}_{i=1}^n\big) &=& \frac{1}{\alpha -1}  \ln \left( \int_{X} \prod_{i=1}^n
 p_i  ^\frac{\alpha}{n} q_i ^\frac{1-\alpha}{n} 
 d\mu \right)\\ &=&\frac{1}{\alpha -1}  \ln \big[D_{  (t^{\alpha }, t^{\alpha }, \cdots, t^{\alpha }) }\big(\vec{\bP}, \vec{\bQ}) \big].
\end{eqnarray*}
The case $\alpha_i=\frac{1}{2}$, for all $i=1, 2, \cdots, n$,  gives the  {\em mixed Bhattcharyya  coefficient} or {\em mixed Bhattcharyya distance} of $(\vec{\bP}, \vec{\bQ})$,  
 \begin{eqnarray*}
D_{  \big(\sqrt{t}, \sqrt{t}, \cdots, \sqrt{t}\big) }\big(\vec{\bP}, \vec{\bQ}) = \int_{X} \prod_{i=1}^n
 p_i  ^\frac{1}{2n} q_i ^\frac{1}{2n} 
 d\mu.
\end{eqnarray*} This integral is related to the Matuista's affinity (see Introduction). For more information on the corresponding $f$-divergences we refer to e.g. \cite{LieseVajda2006}. 

\section {Applications to convex geometry.}
 An important application of the mixed $f$-divergence arises in the theory of convex bodies. A convex body  $K$ in $\mathbb{R}^n$    is a convex, compact subset of $\mathbb{R}^n$ with non-empty interior. We write $\mathcal{K}_0$ for the set of all convex bodies in $\mathbb{R}^n$ with the origin in the interior. We use $|K|$ to denote the volume of $K$ and $|\partial K|$ to denote the surface area of $\partial K$, the boundary of $K$.   We write  $B^n_2$ for the   Euclidean  unit ball  in $\mathbb{R}^n$  and $S^{n-1}$ for the unit sphere in $\mathbb{R}^n$.   The usual inner product in $\mathbb{R}^n$ is denoted by $\langle \cdot, \cdot\rangle$.  
\par
For $K\in\mathcal{K}_0$, 
the {\em polar body $K^\circ$} of $K$ is defined by $$K^\circ=\{y\in \mathbb{R}^n: \langle x, y\rangle \leq 1, \ \ \ \forall x\in K\}.$$   The {\em support function} of $K$, $h_K: S^{n-1}\rightarrow \mathbb{R}^+$,  is $h_K(u) =\max_{x\in K}\langle x, u \rangle$.  For $x \in \partial K$,   $\kappa_K(x)$ is the (generalized) {\em Gaussian curvature}  at $x$. Then, for a convex body $K$ of class of  $C^2_+$, i.e., whose boundary is $C^2$ with strictly positive Gauss curvature everywhere, the  
 {\it curvature function} $f_K(u): S^{n-1}\rightarrow \bbR$ is defined by  $f_K(u) = \frac{1}{\kappa_K(x)}$,  where $x \in \partial K$ is such that the outer normal vector to $\partial K$ at $x$ is $u$. We refer to \cite{Ga, Sch} for more details on convex bodies.

\subsection{General mixed affine surface areas.}
Here, we link the mixed $f$-divergence with general mixed affine surface areas for  convex bodies. 
We let $X=S^{n-1}$ be the unit sphere in $\mathbb{R}^n$ and $\mu$ be the  spherical measure $\sigma$. Let $K_1, \dots , K_n \in \cK_0$ be convex bodies of class of $C^2_+$, and $\xi \in S^{n-1}$. 
For $1 \leq i \leq n$, let
\begin{equation*} 
p_{K_i}(\xi)= \frac{1}{ n |K_i^{\circ}| h_{K_i}^n(\xi)} \, , \   \ q_{K_i}(\xi)= \frac{f_{K_i}(\xi) h_{K_i}(\xi) }{n\  |K_i|},
\end{equation*}
and define probability measures on $S^{n-1}$ by
\begin{equation*} 
 P_{K_i}=p_{K_i}   \sigma \ \ \ \text{and}   \ \ \    Q_{K_i}=q_{K_i}   \sigma.
\end{equation*}
Let $f_i: (0, \infty) \rightarrow \mathbb{R}^+$, $ 1 \leq i \leq n$, be convex and/or concave functions.
Then, we define $D_{\vec{\mathbf{f}}}\big((P_{K_1},  \dots, P_{K_n}), (Q_{K_1}, \dots, Q_{K_n})\big)$
by
\begin{eqnarray}
&&D_{\vec{\mathbf{f}}}\big((P_{K_1},  \dots, P_{K_n}), (Q_{K_1}, \dots, Q_{K_n})\big)\nonumber \\ &&\ \ \ =  \int_{S^{n-1}} \prod_{i=1}^n 
\left[f_i\left(\frac{p_{K_i}}{q_{K_i}}\right) q_{K_i}\right]^\frac{1}{n}  d\sigma \nonumber \\ &&\ \ \ =  \int_{S^{n-\!1}} \! \prod_{i=1}^n 
\left[f_i\!\left(\!\frac{|K_i||K_i^\circ|^{-1}}{ f_{K_i}h_{K_i}^{n+1}}\!\! \right)\!  \frac{f_{K_i} h_{K_i}}{n\  |K_i|} \right]^\frac{1}{n} \!\!\! d\sigma.\ \ \ \ \ \ \ \ \label{mixed:bodies1}
\end{eqnarray}
This expression is  closely related to the general mixed $L_{\phi}$ (or $L_{\psi}$) affine surface areas introduced in  \cite{Ye2012}.
The companion expression
\begin{eqnarray}
\!\! &&D_{\vec{\mathbf{f}}}\big((Q_{K_1}, \dots, Q_{K_n}), (P_{K_1}, \dots, P_{K_n})\big)\nonumber \\ &&\ \ \ =  \int_{S^{n-1}} \prod_{i=1}^n \left[f_i\left(\frac{q_{K_i}}{p_{K_i}}\right) p_{K_i} \right]^\frac{1}{n} d\sigma \nonumber \\ && \ \ \ =\! \int_{S^{n-1}}\! \prod_{i=1}^n 
\left[f_i\!\left(\!\frac{ f_{K_i}h_{K_i}^{n+1}}{|K_i| |K_i^\circ|^{-1}}\!\! \right)  \frac{|K_i^\circ|^{-1}}{n   h_{K_i}^n }\! \right]^\frac{1}{n}\!\!\!  d\sigma,\ \ \ \ \ \ \ \ \label{mixed:bodies2} 
\end{eqnarray} 
is closely related to the general mixed $L_{\phi}^*$ (or $L_{\psi}^*$) affine surface areas introduced in 
\cite{Ye2012}. One can easily obtain that both formulas (\ref{mixed:bodies1}) and (\ref{mixed:bodies2}) are affine invariant. For instance, for all linear transform $T$ with the absolute value of its determinant equal to $1$, one has $D_{\vec{\mathbf{f}}}\big((P_{TK_1},  \dots, P_{TK_n}), (Q_{TK_1}, \dots, Q_{TK_n})\big)$ equal to $D_{\vec{\mathbf{f}}}\big((P_{K_1},  \dots, P_{K_n}), (Q_{K_1}, \dots, Q_{K_n})\big).$ 
\vskip 2mm
The ``principle for change of order" (Proposition \ref{principle}) implies that
$D_{\vec{\mathbf{f}}}\big((P_{K_1}, \dots, P_{K_n}), (Q_{K_1},\dots,Q_{K_n})\big)$ is identical to $
D_{\vec{\mathbf{f}}^*}\big((Q_{K_1}, \dots, Q_{K_n}), (P_{K_1}, \dots, P_{K_n})\big).$ 

\vskip 2mm 
 When all $K_i$ are centrally symmetric Euclidean balls, i.e., for all $i=1, \dots,n$,  $K_i=r_iB_2^n$ for some $r_i>0$, then for all $\xi\in S^{n-1}$,  $h_{K_i}(\xi)=r_i$, and $f_{K_i}(\xi)=r_i^{n-1}$. This implies that $p_{K_i}=q_{K_i}=\frac{1}{n|B^n_2|}$ for all $1\leq i\leq n$, and hence  $$D_{\vec{\mathbf{f}}}\big(\!(P_{r_1B_2^n}, \dots, P_{r_nB_2^n}), (Q_{r_1B_2^n}, \dots, Q_{r_nB_2^n})\!\big)\!\!=\! \prod_{i=1}^n [f_i(1)]^{\frac{1}{n}}\! . $$ The isoperimetric  inequality proved in \cite{Ye2012} (Theorem 3.2) says that under certain conditions on $f_i$ and $K_i$, $$D_{\vec{\mathbf{f}}}\big((Q_{K_1}, \dots, Q_{K_n}), (P_{K_1}, \dots, P_{K_n})\big)\leq \prod_{i=1}^n [f_i(1)]^{1/n},$$  and  the maximum is obtained when all $K_i$ are centrally symmetric Euclidean balls. This inequality will be extended to a more general setting for the mixed $f$-divergence for measures.

\subsection{The $f$-dissimilarity for multiple convex bodies.}
The above connection between general mixed affine surface areas and the mixed $f$-divergence can be further extended to the  $\bff$-dissimilarity for multiple convex bodies. Let $\bff:\bbR^l\rightarrow \bbR$ be a convex function. We consider the measure space $(S^{n-1}, \sigma)$. For a convex body $K\in \cK_0$ of class of $C^2_+$,  $P_K$ is a probability measure associated with $K$. We denote by $p_K$ the density function of $P_K$ with respect to $\sigma$. Likewise,  for  convex bodies $K_i\in \cK_0, 1\leq i\leq l$,  of  class of $C^2_+$, we let $P_{K_i}$ be measures associated with $K_i$ whose Radon-Nikodym derivatives with respect to  the spherical measure $\sigma$ are $p_{K_i}$. Then the $\bff$-dissimilarity of $P_{K_i}$ with reference probability measure $P_K$ is defined as $$D_{\bff}(P_{K_1}, \cdots, P_{K_l}; P_K)=\int_{S^{n-1}} \bff\left(\frac{p_{K_1}}{p_K}, \cdots, \frac{p_{K_l}}{p_K}\right) p_K\,d\sigma.$$ We will also use the notation $D_{\bff}(p_{K_1}, \cdots, p_{K_l}; p_K)$ for  $D_{\bff}(P_{K_1}, \cdots, P_{K_l}; P_K)$
and  $D_{f}(p,q)$ for  $D_{f}(P, Q)$.
\par
Aside from the general (mixed) affine surface areas, many other important objects in convex geometry are special cases of the $f$-dissimilarities. We now give another example.  Let $K=B_2^n$, and $P_K=\frac{\sigma}{|\partial B^n_2|}$.  Let $K_1\in \cK_0$  be of class of $C^2_+$.
Then, using  the convex function $f(x)=x$,  the surface area of $\partial K_1$ is $$\frac{|\partial K_1|}{|\partial B^n_2|}=\frac{1}{|\partial B^n_2|}\int_{S^{n-1}} f_{K_1}(u)\,d\sigma=D_{f}\left(f_{K_1},  \frac{1}{|\partial B^n_2|}\right).$$ Similarly, using the convex function $f(x)=x^{\frac{n+1}{n}}$, $x \geq 0$, one can also write $|\partial K_1|$ as $$\frac{|\partial K_1|}{|\partial B^n_2|} =D_{f}\left([f_{K_1}]^{\frac{n}{n+1}}, \frac{1}{|\partial B^n_2|}\right).$$  By Jensen's  inequality (see also inequality (\ref{Iso:type:1})), one has $$\frac{|\partial K_1|}{|\partial B^n_2|}\geq \left(\frac{as(K_1)}{n|B^n_2|}\right)^{\frac{n+1}{n}} = \left(\frac{as(K_1)}{as(B^n_2)}\right)^{\frac{n+1}{n}},$$ which compares the surface area and the  affine surface area (e.g. \cite{Bl1, lei, Lu1, SW1990}) $$as(K_1)=\int_{S^{n-1}} [f_{K_1}(u)]^{\frac{n}{n+1}}\,d\sigma.$$ 
Note that $f(x)=x^{\frac{n+1}{n}}$ is strictly convex.  Thus equality holds in the above inequality if and only if $f_{K_1}(u)\equiv C$, with $C>0$ a constant, which happens if and only if $K_1$ is a Euclidean ball.

\section{Inequalities.}
 
 The classical Alexandrov-Fenchel
inequality for  mixed volumes of convex bodies is a fundamental result in (convex) geometry. 
A general version of this inequality  for  {\em mixed volumes} of convex bodies $K_1, \dots, K_n$ in $\mathbb{R}^n$  (see \cite{ Ale1937, Bus1958,
Sch}) can be written as, for all integer $m$ s.t. $ 1\leq m\leq n $
$$
\prod_{i=0}^{m-1}\!\!V(K_1,\cdots\!, K_{n-m}, \underbrace{K_{n-i},
\cdots\!, K_{n-i}}_m)\!\leq\! V^m(K_1, \cdots\!, K_n). $$  
We refer to e.g. \cite{Sch} for the definition of the mixed volume $V(K_1,\cdots, K_{n})$ of  the bodies $K_1, \dots, K_n$.

\par
Alexandrov-Fenchel type inequality for 
 the (mixed) affine surface areas can be found in
\cite{Lut1987, Lu1, WernerYe2010, Ye2012}. Now we
prove  an Alexandrov-Fenchel type inequality for the 
mixed $f$-divergence for measures.
\vskip 2mm
Following \cite{HardyLittlewoodPolya}, we say that two functions $f$ and $g$ are {\em effectively proportional}  if there are 
constants $a$ and $b$, not both zero, such that $af=bg$. Functions $f_1, \dots, f_m$ are effectively proportional if every pair $(f_i, f_j), 1\leq i, j\leq m$ is  effectively proportional.  A null function is effectively proportional to any function.
These notions will be used in the next theorems. 
\par
For a measure space 
$(X, \mu)$ and probability densities $p_i $ and
$q_i$, $1 \leq i \leq n$, we put
\begin{equation}\label{g0}
g_0(u)= \prod
_{i=1}^{n-m} \left[f_i\left(\frac{p_{i}}{q_{i}}\right)
q_{i}\right]^\frac{1}{n},
\end{equation} and for $j=0, \cdots, m-1$, 
\begin{equation}\label{gi}
g_{j+1}(u)=
\left[f_{n-j}\left(\frac{p_{n-j}}{q_{n-j}}\right)
q_{n-j}\right]^\frac{1}{n}.
\end{equation} For a vector $\vec{p}$, we denote by $\vec{p}^{\ n,k}$ the following vector $$\vec{p}^{\ n, k}=(p_1, \cdots, p_{n-m}, \underbrace{p_k, \cdots, p_k}_m), \ \ \ k>n-m.$$
\bt \label{inequality:mixed:f:divergence} Let $(X, \mu)$ be a
finite measure space. For $1 \leq i \leq n$, let $P_i$ and
$Q_i$ be probability measures on $(X, \mu)$ with density functions $p_i$ and $q_i$ respectively almost everywhere w.r.t.\!\! $\mu$. Let $f_i: (0,
\infty) \rightarrow \mathbb{R}^+$, $ 1 \leq i \leq n$, be convex
functions. Then, for $1 \leq m \leq n$,\begin{small}
$$ \big[\dfpq\big]^m \leq \prod_{k =n-m+1}^ {n} D_{\vec{f}^{n, k}}\big(\vec{\bP}^{n, k},\vec{\bQ}^{n, k}\big).$$
\end{small} 
\noindent Equality holds if and only if one of the functions $g_0^\frac{1}{m} g_{i}$, $1 \leq i \leq m$,  is null or all are effectively proportional $\mu$-a.e. 
\par
\noindent If $m=n$,
\begin{eqnarray*} [\dfpq]^n
\leq \prod_{i=1}^n D_{f_i}(P_i, Q_i), 
\end{eqnarray*} 
with equality if and only if one of the functions  $f_{j}\left(\frac{p_{j}}{q_{j}}\right)
q_{j}$, $0 \leq j \leq n$, is null or all  are effectively proportional $\mu$-a.e.
\et

 \noindent {\bf Remarks.}  
(i) In particular, 
equality holds in Theorem
\ref{inequality:mixed:f:divergence} if all $(P_i, Q_i)$ coincide, and $f_i=\lambda_i f$ 
for some convex positive function $f$ and $\lambda_i\geq 0$, $i=1, 2, \cdots, n$.
\par
\noindent
(ii) 
 Theorem
\ref{inequality:mixed:f:divergence} still holds true if  the functions $f_i$ are
concave.

\vskip 2mm \noindent \begin{IEEEproof}[Proof of Theorem 
\ref{inequality:mixed:f:divergence}] We let $g_0$ and $g_{j+1}$, $j=0, \cdots, m-1$
as in (\ref{g0}) and (\ref{gi}). 
By H\"{o}lder's inequality (see
\cite{HardyLittlewoodPolya})\begin{small}
\begin{eqnarray*}
[\dfpq]^m&=&\left(\int
_{X}g_0(u) g_1(u) \cdots g_{m}(u)\,d\mu\right)^m\\ &=& \bigg(\int
_{X}\prod_{j=0}^{m-1}\left[g_0(u) g_{j+1}(u)^m\right]^{\frac{1}{m}}\,d\mu\bigg)^m\\ &\leq& \prod _{j=0}^{m-1} \left(\int _{X} g_0(u)
g_{j+1}^m(u)\,d\mu\right)\\ &=&\prod_{k =n-m+1}^ {n} D_{\vec{f}^{n, k}}\big(\vec{\bP}^{n, k},\vec{\bQ}^{n, k}\big).
\end{eqnarray*}
\end{small} By e.g.
\cite{HardyLittlewoodPolya}, equality holds in H\"{o}lder's inequality, if and only if 
one of the functions $g_0^\frac{1}{m} g_{i}$, $1 \leq i \leq m$,  is null or all are effectively proportional $\mu$-a.e..

\vskip 2mm 
In particular, this is the case, if for all $i=1,  \cdots, n$, $(P_i, Q_i)=(P, Q)$ and $f_i=\lambda_i f$
for some convex function $f$ and $\lambda_i\geq 0$.  \end{IEEEproof}

\vskip 3mm 
Let $f:(0,\infty) \rightarrow \mathbb{R}^+$ be a convex function. By Jensen's inequality,
 \begin{equation}\label{Iso:type:1} D_f(P, Q)\!=\!\int_X
\!f\!\left(\!\frac{p}{q}\!\right)q\,d\mu\geq f\!\left(\!\int_X
\!p\,d\mu\!\right)=f(1),\end{equation} for all pairs of probability
measures $(P, Q)$ on $(X, \mu)$ with nonzero density functions $p$ and $q$ respectively almost everywhere w.r.t.\!\! $\mu$.  When $f$ is linear, equality holds trivially in (\ref {Iso:type:1}) . When $f$ is strictly convex, equality holds
true if and only if $p=q$ almost everywhere with respect to the measure $\mu$.
If $f$ is a concave function, again by
Jensen's inequality,  \begin{equation}\label{Iso:type:2}
D_f(P, Q)\!=\!\int_X\! f\!\left(\!\frac{p}{q}\!\right)q\,d\mu\leq
f\!\left(\!\int_X\! p\,d\mu\!\right)=f(1),\end{equation} for all pairs of
probability measures $(P, Q)$.  Again, when $f$ is linear, equality holds trivially. When $f$ is strictly concave,
equality holds true if and only if $p=q$ almost everywhere with respect to the measure $\mu$.

\vskip 2mm For the mixed $f$-divergence with concave functions,
one has the following result.

\bt  Let $(X, \mu)$ be a finite measure space. For all $1 \leq i \leq n$, let
$P_i$ and $Q_i$ be probability measures on $X$ whose density functions $p_i$ and $q_i$ are nonzero almost everywhere w.r.t.\!\! $\mu$.
Let $f_i: (0, \infty) \rightarrow \mathbb{R}^+$, $ 1 \leq i \leq n$,
be concave functions. Then
\begin{eqnarray}\label{inequality:mixed:f:2} [\dfpq]^n
\leq \prod_{i=1}^n D_{f_i}(P_i, Q_i) \leq   \prod_{i=1}^nf_i(1).
\end{eqnarray} 

\noindent If in addition, all $f_i$ are strictly concave, equality holds if and only if there is a probability density $p$ such that
 $$p_i=q_i=p, \ \ 1\leq i\leq n,  $$ almost everywhere with respect to the measure $\mu$.
\et

\vskip 2mm 
\noindent \begin{IEEEproof} Theorem
\ref{inequality:mixed:f:divergence} and the remark after imply that for all concave
functions $f_i$,
\begin{eqnarray*}[\dfpq]^n \leq \prod_{i=1}^n D_{f_i}(P_i, Q_i)\leq  \prod_{i=1}^n f_i(1),
\end{eqnarray*} where the second inequality follows from
inequality (\ref{Iso:type:2}) and $f_i\geq 0$. 
\par
Suppose now that for all $i$, $p_i=q_i=p$, $\mu$-a.e., where $p$ is a fixed probability density. 
Then equality holds trivially in (\ref{inequality:mixed:f:2}).

Conversely, suppose that equality holds in (\ref{inequality:mixed:f:2}). Then, in particular, equality holds in Jensen's inequality which, as noted above, happens if and only if $p_i=q_i$ for all $i$. Thus,    $$\dfpq =\left(\prod_{i=1}^n [f_i(1)]^{1/n}\right) \int _{X} q_1^{1/n} \dots \,q_n^{1/n}d\mu.$$ Note also that if all $f_i: (0, \infty)\rightarrow \mathbb{R}^+$ are strictly concave, $f_{i}(1)\neq 0$ for all $1\leq i\leq n$.  Equality characterization in H\"older's inequality  implies that all $q_i$ are effectively proportional $\mu$-a.e. As all $q_i$ are probability measures,  they are all equal (almost everywhere w.r.t.\!\! $\mu$) to a probability measure with density function (say) $p$.  \end{IEEEproof}
\vskip 2mm
\noindent
{\bf Remark.}
If $f_i(t)=a_i t +b_i$ are all linear and positive, then equality holds if and only if all $p_i, q_i$ are equal (almost everywhere w.r.t. $\mu$) as convex combinations, i.e., if and only if for all $i, j$ $$
\frac{a_i}{a_i+b_i} p_i + \frac{b_i}{a_i+b_i} q_i = \frac{a_j}{a_j+b_j} p_j + \frac{b_j}{a_j+b_j} q_j,  \hskip 4mm \mu - \text{a.e.}
$$
 
\section{The $i$-th mixed $f$-divergence.}
Let $(X, \mu)$ be a finite measure space. Throughout this section, we assume that the functions  $$f_1, f_2: (0, \infty)\rightarrow \{x\in \mathbb{R}: x>0\},$$ are  convex or concave, and that {$P_1, P_2, Q_1, Q_2$  are probability
measures on $X$ with density functions $p_1, p_2, q_1, q_2$ which are nonzero almost everywhere w.r.t. the measure $\mu$.}  We also write $$\vf=(f_1, f_2), \ \ \vP=(P_1, P_2), \ \ \vQ=(Q_1, Q_2).$$

\begin{definition} Let $i\in \bbR$. The $i$-th mixed
$f$-divergence for $(\vf, \vP, \vQ)$,
denoted by $\dfpqi$,
is defined as \begin{equation} \dfpqi\!=\!\!\!\int_{X}\!\!
\left[f_1\!\left(\frac{p_{1}}{q_{1}}\right)\! q_{1}\right]^\frac{i}{n}\!\!
\left[f_2\!\left(\frac{p_{2}}{q_{2}}\right)\!
q_{2}\right]^\frac{n-i}{n} \!\!\!\!\! d \mu.
\label{i:mixed:phi}\end{equation}
\end{definition} 
\vskip 2mm
\noindent
{\bf Remarks.} 
Note that  the  $i$-th mixed
$f$-divergence is  defined for any combination of convexity and concavity of $f_1$ and $f_2$,  namely,  both $f_1$ and $f_2$ concave, or both $f_1$ and $ f_2$ convex, or one is  convex
the other  is concave.  
\par
\noindent
It is easily checked that $$\dfpqi=D_{(f_2, f_1)}\big((P_2, Q_2), (P_1,
Q_1); n-i\big).$$
If $0\leq i\leq n$ is an integer, then the triple  $(f_1,
P_1, Q_1)$ appears $i$-times while the triple $(f_2, P_2, Q_2)$ appears
$(n-i)$ times in $\dfpqi$. 
\par
\noindent
For $i=0$,
$$\dfpqi=D_{f_2}(P_2,
Q_2), $$ and for $i=n$, $$\dfpqi=D_{f_1}(P_1, Q_1).$$ 
\par
Another special case is when
$P_2=Q_2=\mu$ almost everywhere and $\mu$ is also a probability measure. Then such an $i$-th mixed $f$-divergence, denoted by
$D\big((f_1, P_1, Q_1), i; f_2\big)$, has the  form
\begin{equation*}
D\big((f_1, P_1, Q_1), i; f_2\big)=[f_2(1)]^{1-i/n}
\int_{X} \left[f_1\left(\frac{p_{1}}{q_{1}}\right)
q_{1}\right]^\frac{i}{n} d \mu.
\end{equation*}
 \vskip 2mm
\noindent
{\bf Examples and Applications.}
\par
\noindent (i) For $f(t)=|t-1|$, we get the {\em $i$-th mixed total variation} $$D_{(f, f)}\big(\vP, \vQ; i\big) =\int_X |p_1-	q_1|^{\frac{i}{n}} |p_2-	q_2|^{\frac{n-i}{n}}\,d\mu.$$  
\par
\noindent
(ii) For $f_1(t)= f_2(t)= [t \ln t]_+$, we get the (modified)  {\em $i$-th mixed relative entropy} or {\em
 $i$-th mixed Kullback Leibler divergence}
 \begin{eqnarray*}
D_{KL}\big(\vP, \vQ; i\big)  = \int_{X} 
\left[ p_1 \ln\left(\frac{p_{1}}{q_{1}}\right) 
\right]_+^\frac{i}{n}  \left[ p_2  \ln\left(\frac{p_{2}}{q_{2}}\right) 
\right]_+^\frac{n-i}{n}  d\mu.
\end{eqnarray*}
\par
\noindent
(iii) For the convex or concave functions $f_{\alpha_j}(t) = t^{\alpha_j}$, $j=1,2$, we get the {\em $i$-th mixed Hellinger integrals}
\begin{eqnarray*}
D_{(f_{\alpha_1}, f_{\alpha_2})}\big( \vP, \vQ; i\big) = \int_{X} 
\left( p_1  ^{\alpha_1} q_1 ^{1-\alpha_1} \right) ^\frac{i}{n}   \left( p_2  ^{\alpha_2} q_2 ^{1-\alpha_2} \right) ^\frac{n-i}{n} 
 d\mu.
\end{eqnarray*}
In particular, for $\alpha_j= \alpha$, for  $j=1,2$, 
\begin{eqnarray*}
D_{(f_{\alpha}, f_{\alpha})}\big( \vP, \vQ; i\big)  = \int_{X} 
\left( p_1  ^{\alpha} q_1 ^{1-\alpha} \right) ^\frac{i}{n}   \left( p_2  ^{\alpha} q_2 ^{1-\alpha} \right) ^\frac{n-i}{n} 
 d\mu.
\end{eqnarray*}
This integral  can be used to define the {\em $i$-th mixed $\alpha$-R\'enyi divergence }
 \begin{eqnarray*}
D_{\alpha}\big(\vP, \vQ; i \big) = \frac{1}{\alpha -1}  \ln \left[D_{(f_{\alpha}, f_{\alpha})}\big( \vP, \vQ; i\big)\right] .
\end{eqnarray*}
The case $\alpha_i=\frac{1}{2}$ for all $i$ gives 
\begin{eqnarray*}
D_{(\sqrt{t}, \sqrt{t})}\big( \vP, \vQ; i\big)  = \int_{X} 
\left( p_1  q_1  \right) ^\frac{i}{2n}   \left( p_2   q_2 \right) ^\frac{n-i}{2n} 
 d\mu,
\end{eqnarray*}
the {\em $i$-th mixed Bhattcharyya  coefficient} or 
{\em $i$-th mixed Bhattcharyya distance} of the $p_i$ and $q_i$.
\par
\noindent
(iv) Important  applications are again in the theory of convex bodies.  As in section 2,
 let $K_1, K_2\in \cK_0$ be convex bodies with positive curvature function. For $l=1,2$, let
\begin{equation*}
p_{K_l}(\xi)= \frac{1}{ n |K_l^{\circ}| h_{K_l}^n(\xi)} \, , \   \ q_{K_l}(\xi)= \frac{f_{K_l}(\xi) h_{K_l}(\xi) }{n\  |K_l|},
\end{equation*}
and define  probability measures on $S^{n-1}$ by
\begin{equation*}
 P_{K_l}=p_{K_l}  \sigma \ \ \ \text{and}   \ \ \    Q_{K_l}=q_{K_l}   \sigma.
\end{equation*}
Let $f_l: (0, \infty) \rightarrow \mathbb{R}$, $ l=1, 2$, be positive convex functions.
Then, we define the {\em $i$-th mixed $f$-divergence} for the convex bodies $K_1$ and $ K_2$ 
by
\begin{eqnarray*}
&&D_{\vf}\big( (P_{K_1}, P_{K_2}), (Q_{K_1}, Q_{K_2}); i\big)\nonumber \\ &&\ \ =  \int_{S^{n-1}} \left[f_1\left(\frac{|K_1^\circ|^{-1}|K_1|}{ f_{K_1}h_{K_1}^{n+1}}\right)  \frac{f_{K_1} h_{K_1}}{n  |K_1|} \right]^\frac{i}{n} \\ && \ \ \ \ \ \ \ \ \times \left[f_2\left(\frac{ |K_2^\circ|^{-1}|K_2| }{f_{K_2}h_{K_2}^{n+1}}\right)  \frac{f_{K_2} h_{K_2}}{n |K_2|} \right]^\frac{n-i}{n}\!\!\! d\sigma.
\end{eqnarray*}
This expression  is closely related to the general $i$-th mixed $L_{\phi}$ (or $L_{\psi}$) affine surface areas introduced in \cite{Ye2012}. Similarly, 
\begin{eqnarray*}
&&D_{\vf}\big( (Q_{K_1}, Q_{K_2}), (P_{K_1}, P_{K_2}); i\big)\nonumber \\ &&\ \ =  \int_{S^{n-1}}\!\! \left[\!f_1\!\!\left(\frac{ f_{K_1}h_{K_1}^{n+1}}{|K_1^\circ|^{-1}|K_1|}\right)   \frac{1}{n |K_1^\circ|h_{K_1}^n}\! \right]^\frac{i}{n}\\ &&\ \ \ \ \ \ \ \ \times \left[\! f_2\!\!\left(\frac{f_{K_2}h_{K_2}^{n+1}}{|K_2^\circ|^{-1}|K_2| }\right)  \frac{1}{n |K_2^\circ|h_{K_2}^n}\! \right]^\frac{n-i}{n}\!\!\!\! d\sigma, 
\end{eqnarray*}
which is closely related to the general $i$th mixed $L_{\phi}^*$ (or $L_{\psi}^*$) affine surface areas introduced in \cite{Ye2012}.

\vskip 2mm  The following result holds for all possible combinations of convexity and concavity 
of $f_1$ and $ f_2$.
\vskip 2mm
\begin{proposition} \label{Monotone:1} Let $\vf, \vP, \vQ$ be as above. 
If $j\leq i\leq k$ or $k\leq i\leq j$, then
\begin{eqnarray*} \dfpqi\leq  \bigg[D_{\vf}\big(\vP, \vQ; j\big)\bigg]^{\frac{k-i}{k-j}}\times \bigg[D_{\vf}\big(\vP, \vQ; k\big)\bigg]^{\frac{i-j}{k-j}}.\end{eqnarray*}
Equality holds trivially if $i=k$ or $i=j$. Otherwise, 
equality holds if and only if one of the functions $f_i\left(\frac{p_i}{q_i}\right) q_i$, $i=1,2$,  is null,  or $f_1\left(\frac{p_1}{q_1}\right) q_1$ and $f_2\left(\frac{p_2}{q_2}\right) q_2$ are effectively proportional $\mu$-a.e.
In particular, this holds if $(P_1, Q_1)=(P_2, Q_2)$ and $f_1=\l f_2$ for
some $\l>0$.
\end{proposition}

\vskip 2mm \noindent \begin{IEEEproof} By formula (\ref{i:mixed:phi}),
one has \begin{small}\begin{eqnarray*} \dfpqi\!\!&\!\!=\!\!&\!\!\int_{X}
\left[f_1\left(\frac{p_{1}}{q_{1}}\right) q_{1}\right]^\frac{i}{n}
\left[f_2\left(\frac{p_{2}}{q_{2}}\right)
q_{2}\right]^\frac{n-i}{n} d \mu\\ \!\!&\!\!=\!\!&\!\!\int
_{X}\!\!\left\{\!\left[\! f_1\left(\!\frac{p_{1}}{q_{1}}\!\right)
q_{1}\!\right]^\frac{j}{n}\!\! \left[\!f_2\!\left(\!\frac{p_{2}}{q_{2}}\!\right)
q_{2}\!\right]^\frac{n-j}{n}\!\right\}^{\frac{k-i}{k-j}}\\ && \times  \left\{\!\left[\!f_1\left(\!\frac{p_{1}}{q_{1}}\!\right)
q_{1}\!\right]^\frac{k}{n}\!\! \left[\!f_2\left(\!\frac{p_{2}}{q_{2}}\!\right)
q_{2}\!\right]^\frac{n-k}{n}\!\right\}^{\frac{i-j}{k-j}}\!\!\!\! d\mu \\ \!\!&\!\!\leq\!\!&\!\! \bigg[D_{\vf}\big(\vP, \vQ; j\big)\bigg]^{\frac{k-i}{k-j}}\times \bigg[D_{\vf}\big(\vP, \vQ; k\big)\bigg]^{\frac{i-j}{k-j}},
\end{eqnarray*}\end{small}
where the last inequality follows from H\"{o}lder's  inequality and
formula (\ref{i:mixed:phi}). 
The equality characterization follows from the one in H\"{o}lder inequality.
In particular, if $(P_1, Q_1)=(P_2, Q_2)$,
and $f_1=\l f_2$ for some $\l>0$,  equality holds.  \end{IEEEproof}
\vskip 3mm
\bc \label{KOR}Let $f_1$ and $ f_2$ be positive, concave functions on $(0, \infty)$. Then
for all $\vP, \vQ$ and  for all
$0\leq i\leq n$,
\begin{equation*} \big[\dfpqi\big]^n\leq [f_1(1)]^i
[f_2(1)]^{n-i}.\end{equation*}
If in addition, $f_1$ and $f_2$ are strictly concave, equality holds if and only if 
$p_1=p_2=q_1=q_2$ $\mu$-a.e.
\ec
\vskip 2mm 
\noindent \begin{IEEEproof} Let $j=0$ and  $k=n$ in Proposition
\ref{Monotone:1}.  Then for all $0\leq i\leq n$,
\begin{eqnarray*}\label{i:mixed:phi:1}
\big[\dfpqi\big]^n &\leq&
[D_{f_1}(P_1, Q_1)]^{i}[D_{f_2}(P_2, Q_2)]^{n-i}\\&\leq& [f_1(1)]^i [f_2(1)]^{n-i},\nonumber\end{eqnarray*} where the
last inequality follows from inequality (\ref{Iso:type:2}).
\par
To have equality, the above inequalities should be equalities. By Proposition \ref{Monotone:1}, one has then that  $f_1\left(\frac{p_1}{q_1}\right) q_1$ and $f_2\left(\frac{p_2}{q_2}\right) q_2$ are effectively proportional $\mu$-a.e. As both,  $f_1$ and $f_2$,  are strictly concave,  Jensen's inequality requires that $p_1=q_1$ and $p_2=q_2$ $\mu$-a.e. Therefore, equality holds if and only if $f_1(1)q_1$ and $f_2(1)q_2$ are effectively proportional $\mu$-a.e. As both,  $f_1(1)$ and $f_2(1)$,  are not zero,  equality holds iff $p_1=p_2=q_1=q_2$ $\mu$-a.e.    \end{IEEEproof}

\vskip 2mm
\noindent
{\bf Remark.}
If $f_1(t) =a_1t+b_1$ and $f_2(t)=a_2t+b_2$ are both linear,  equality holds in Corollary \ref{KOR}  if and only if
$p_i, q_i$, $i=1,2$, are equal as convex combinations,  i.e., if and only if 
$$
\frac{a_1}{a_1+b_1} p_1 + \frac{b_1}{a_1+b_1} q_1 = \frac{a_2}{a_2+b_2} p_2 + \frac{b_2}{a_2+b_2} q_2,  \hskip 4mm \mu - \text{a.e.}
$$
\vskip 2mm

This proof can be used to establish the following result for
$D\big((f_1, P_1, Q_1), i; f_2\big)$.

\bc Let $(X, \mu)$ be a probability space. Let $f_1$ be a positive concave
function on $(0, \infty)$. Then for all $P_1, Q_1$, for
all (concave or convex) positive functions $f_2$, and for all $0\leq
i\leq n$,
\begin{equation*} \big[D\big((f_1, P_1, Q_1), i;
f_2\big)\big]^n\leq [f_1(1)]^i [f_2(1)]^{n-i}.\end{equation*} 
If $f_1$ is strictly concave, equality holds  if and only if $P_1=Q_1=\mu$.
When $f_1(t) =at+b$ is linear, equality holds if and only if ${ap_1+bq_1}={a+b}$ $\mu$-a.e.\ec

\vskip 2mm

\bc \label{KOR1} Let $f_1$ be a positive convex function and $f_2$ be a positive concave function on $(0,
\infty)$. Then, for all $\vP, \vQ$, and  for all $k\geq n$,
\begin{equation*} \big[D_{\vf}\big(\vP, \vQ;  k\big)\big]^n\geq [f_1(1)]^k
[f_2(1)]^{n-k}.\end{equation*} 
If in addition, $f_1$ is strictly convex and $f_2$ is strictly concave, equality holds
if and only if $p_1=p_2=q_1=q_2$ $\mu$-a.e.
\ec 
\vskip 2mm 
\noindent 
\begin{IEEEproof}  On the right hand side of Proposition \ref{Monotone:1}, let $i=n$ and $ j=0$. Let  $k\geq n$. Then
\begin{eqnarray*}
\big[D_{\vf}\big(\vP, \vQ;  k\big)\big]^n &\geq&  [D_{f_1}(P_1, Q_1 )]^{k}[D_{f_2}(P_2, Q_2
)]^{n-k}\\ &\geq& [f_1(1)]^k [f_2(1)]^{n-k}.\end{eqnarray*} Here, the
last inequality follows from inequalities (\ref{Iso:type:1}), (\ref{Iso:type:2}) and $k\geq n$. 
To have equality, the above inequalities should be equalities. By Proposition \ref{Monotone:1}, one has then that  $f_1\left(\frac{p_1}{q_1}\right) q_1$ and $f_2\left(\frac{p_2}{q_2}\right) q_2$ are effectively proportional $\mu$-a.e. As $f_1$ is strictly convex and $f_2$ is strictly concave,  Jensen's inequality implies that $p_1=q_1$ and $p_2=q_2$ $\mu$-a.e. Therefore, as 
both $f_1(1)$ and $f_2(1)$ are not zero, equality holds if and only if $p_1=p_2=q_1=q_2$ $\mu$-a.e.  
\end{IEEEproof} 
\vskip 2mm
\noindent
{\bf Remark.}
If $f_1(t) =a_1t+b_1$ and $f_2(t)=a_2t+b_2$ are both linear,  equality holds in Corollary \ref{KOR1}  if and only if
$p_i, q_i$, $i=1,2$, are equal $\mu$-a.e. as convex combinations, i.e., if and only if
$$
\frac{a_1}{a_1+b_1} p_1 + \frac{b_1}{a_1+b_1} q_1 = \frac{a_2}{a_2+b_2} p_2 + \frac{b_2}{a_2+b_2} q_2,  \hskip 4mm \mu - \text{a.e.}
$$

This proof can be used to establish the following results for
$D\big((f_1, P_1, Q_1), k; f_2\big)$. 
\vskip 3mm
\bc Let $(X, \mu)$ be a probability space. Let $f_1$ be a positive convex
function on $(0, \infty)$. Then for all $P_1, Q_1$,  for
all (positive concave or convex) functions $f_2$, and for all $k\geq n$,
\begin{equation*} \big[D\big((f_1, P_1, Q_1), k;
f_2\big)\big]^n\geq [f_1(1)]^k [f_2(1)]^{n-k}. \end{equation*}If $f_1$ is strictly convex, equality holds if and only if $P_1=Q_1=\mu$.
When $f_1(t) =at+b$ is linear, equality holds if and only if ${ap_1+bq_1}={a+b}$ $\mu$-a.e.

\ec
\vskip 3mm
\bc 
Let $f_1$ be a positive concave function and $f_2$ be a positive convex function on $(0, \infty)$. Then for
all $\vP, \vQ$,  and  for all $k\leq 0$,
\begin{equation*} \big[D_{\vf}(\vP, \vQ; k)\big]^n\geq [f_1(1)]^k [f_2(1)]^{n-k}.\end{equation*} 
If in addition, $f_1$ is strictly concave and $f_2$ is strictly convex, equality holds iff 
$p_1=p_2=q_1=q_2$ $\mu$-a.e.
\ec
\vskip 2mm \noindent \begin{IEEEproof}  Let $i=0$ and $j=n$ in Proposition
\ref{Monotone:1}. Then
\begin{eqnarray*}
\big[D_{\vf}\big(\vP, \vQ; k)\big]^n  &\geq&  [D_{f_1}(P_1, Q_1 )]^{k}[D_{f_2}(P_2, Q_2 )]^{n-k}\\ &\geq& [f_1(1)]^k [f_2(1)]^{n-k}.\end{eqnarray*} Here, the last
inequality follows from inequalities (\ref{Iso:type:1}), (\ref{Iso:type:2}), and $k\leq 0$.

To have equality, the above inequalities should be equalities. By Proposition \ref{Monotone:1}, one has then that  $f_1\left(\frac{p_1}{q_1}\right) q_1$ and $f_2\left(\frac{p_2}{q_2}\right) q_2$ are effectively proportional $\mu$-a.e. As $f_1$ is strictly concave and $f_2$ is strictly convex, Jensen's inequality requires that $p_1=q_1$ and $p_2=q_2$. Therefore, equality holds if and only if $f_1(1)q_1$ and $f_2(1)q_2$ are effectively proportional $\mu$-a.e.  As both $f_1(1)$ and $f_2(1)$ are not zero,  equality holds if and only if $p_1=p_2=q_1=q_2$ $\mu$-a.e.  \end{IEEEproof} 
 
 \vskip 3mm 
This proof can be used to establish the following results for $D\big((f_1, P_1, Q_1), k; f_2\big)$.
 \vskip 2mm
\bc 
Let $f_1$ be a concave function on $(0, \infty)$. Then for
all $P_1, Q_1$, for all (concave or convex) functions
$f_2$, and for all $k\leq 0$,
\begin{equation*} \big[D\big((f_1, P_1, Q_1), k;  f_2\big)\big]^n\geq [f_1(1)]^k [f_2(1)]^{n-k}.\end{equation*}
If $f_1$ is strictly concave, equality holds if and only if $P_1=Q_1=\mu$.
When $f_1(t) =at+b$ is linear, equality holds if and only if ${ap_1+bq_1}={a+b}$ $\mu$-a.e.

\ec



%



\section*{Acknowledgment}

The research of Elisabeth Werner is partially supported by an NSF grant. The research of Deping Ye is supported by an NSERC grant and a
start-up grant from Memorial University of Newfoundland.

\ifCLASSOPTIONcaptionsoff
  \newpage
\fi

\IEEEtriggercmd{\enlargethispage{-5in}}


\bibliographystyle{IEEEtran}
%

%








\end{document}